\documentclass[pdflatex,sn-mathphys-num]{sn-jnl}% Math and Physical Sciences Numbered Reference Style
\usepackage{graphicx}%
\usepackage{multirow}%
\usepackage{amsmath,amssymb,amsfonts}%
\usepackage{amsthm}%
\usepackage{mathrsfs}%
\usepackage[title]{appendix}%
\usepackage{xcolor}%
\usepackage{textcomp}%
\usepackage{xurl}%
\usepackage{manyfoot}%
\usepackage{booktabs}%
\usepackage{algorithm}%
\usepackage{algorithmicx}%
\usepackage{algpseudocode}%
\usepackage{listings}%
\usepackage{lineno}
\theoremstyle{thmstyleone}%
\theoremstyle{thmstyletwo}%

\theoremstyle{thmstylethree}%
\begin{document}

\title[Article Title]{Non-destructive 3D doping imaging of silicon sensors}


\author*{\fnm{Xiangyu} \sur{Xie}\textsuperscript{*}}\email{xiangyu.xie@psi.ch}
\author{\fnm{Anna} \sur{Bergamaschi}} % Supervision
\author{\fnm{Maria} \sur{Carulla}} % Resources
\author{\fnm{Roberto} \sur{Dinapoli}} % ASIC design and review
\author{\fnm{Erik} \sur{Fröjdh}} % Data analysis library
\author{\fnm{Viktoria} \sur{Hinger}} % Review
\author{\fnm{Davide} \sur{Mezza}} % Backside pulsing method
\author{\fnm{Aldo} \sur{Mozzanica}} % ASIC design and review
\author{\fnm{Jonathan} \sur{Mulvey}} % Review and discussion
\author{\fnm{Bernd} \sur{Schmitt}} % Supervision and review
\author{\fnm{Saverio} \sur{Silletta}} % Review
\author{\fnm{Jiaguo} \sur{Zhang}} % Conceptualization

\affil{\orgdiv{Photon Science Detector Group}, \orgname{Paul Scherrer Institute}, \orgaddress{\street{Forschungsstrasse 111}, \city{Villigen}, \postcode{5232}, \country{Switzerland}}}


\abstract{
    Silicon sensors are the foundational detection medium for X-rays and charged particles.
    While their bulk dopant distribution determines device performance, it is conventionally assumed homogeneous because traditional profiling is destructive, spatially restricted, and insensitive at the relevant concentrations.
    Here we introduce a non-destructive 3D doping imaging technique that turns the readout electronics of a charge-integrating hybrid pixel detector into a massively parallelized capacitance-voltage profiler.
    With a few tens of micrometres of 3D resolution over wafer-scale areas at concentrations on the order of $10^{11}\ \text{cm}^{-3}$, we image the bulk doping concentration of operational sensors.
    Macroscopically, we resolve depth-evolving concentric doping rings; microscopically, we uncover scattered doping anomalies that distort local electric fields.
    The rings modulate the depletion voltage, while the anomalies disrupt local charge collection, a previously overlooked cause of pixel yield and performance degradation.
    By bridging manufacturing signatures with microscopic defects, this approach provides a non-destructive framework for sensor characterization and yield optimization.
}

% \keywords{Silicon Sensor, 3D Doping Imaging, Non-Destructive Characterization, Hybrid Pixel Detector}



\maketitle

\section*{Introduction}\label{sec1}

Silicon sensors form the cornerstone of hybrid pixel detectors~\cite{Delpierre_2014, KASTLI2006188, Broennimann_2006}, underpinning groundbreaking discoveries and technological leaps across particle physics~\cite{HpdForHep}, photon science~\cite{HpdForPS}, electron microscopy~\cite{FARUQI2003263}, and medical imaging~\cite{BALLABRIGA2020106271}.
The three-dimensional spatial distribution of dopants within the sensor bulk is a fundamental parameter that determines the electrical properties and reliability of these devices. 
As the development of advanced detectors pushes simultaneously towards macroscopic dimensions~\cite{AKoch_2013} while demanding increasingly finer micrometre-regime granularity~\cite{Bergamaschi02112018}, achieving uniform doping across large sensor areas has become essential for ensuring consistent performance and maximizing manufacturing yield.
However, comprehensive three-dimensional characterization of these doping profiles remains a persistent challenge, often forcing the community to rely on presumed bulk uniformity.

Conventional characterization techniques present dimensional and practical limitations when applied to modern high-resistivity silicon sensors.
Standard macroscopic non-destructive methods, such as four-point probe and Capacitance-Voltage (C-V) profiling~\cite{PBlood_1986}, provide only laterally averaged surface-level properties.
Conversely, high-resolution techniques like Secondary Ion Mass Spectrometry (SIMS) and Spreading Resistance Profiling (SRP) offer excellent vertical precision for bulk characterization, but they essentially yield localized 1D depth tracks.
Extending them to 3D imaging is destructive, time-consuming, and impractical. 
Furthermore, these techniques typically suffer from detection limits no better than $\sim 10^{13}\ \text{cm}^{-3}$ in silicon~\cite{SIMS_Intro, SIMs_sensitivity, SSR_sensitivity}, leaving them insensitive to the low doping concentrations ($\sim 10^{12}\ \text{cm}^{-3}$ or lower) characteristic of high-resistivity sensor bulks. 

To bridge this gap, we present a non-destructive 3D doping imaging technique that turns the pixelated charge-integrating readout electronics into a massively parallelized capacitance-voltage profiling array.
By integrating a backside pulsing scheme~\cite{Mezza_2016} with synchronized bias sweeping, our approach enables \textit{in-situ}, high-resolution mapping of fully assembled detectors.
This method achieves a combination of 3D resolution and large-area coverage not accessible to established profiling techniques: a lateral resolution set by the pixel pitch, down to 25~\textmu m, and a depth resolution of ${\sim}20$~\textmu m, over active areas up to tens of $\mathrm{cm^{2}}$.
Furthermore, it operates in the concentration range characteristic of high-resistivity sensor bulks, a few $10^{11}\ \text{cm}^{-3}$, where it resolves relative variations at the sub-percent level.

Utilizing this technique, we visualize the complex, three-dimensional bulk doping landscape of functional sensors. 
Macroscopically, we resolve concentric doping ring patterns that dynamically evolve with depth and uniquely correlate with the float-zone crystal growth process, introducing long-range spatial fluctuations in the depletion voltage.
At the microscopic scale, we identify doping anomalies widely scattered across the silicon bulk.
These intrinsic material defects, previously obscured by conventional lateral averaging techniques, distort local electric fields and emerge as a critical, yet largely unrecognized, root cause of pixel yield degradation.

\section*{Results}\label{sec2}

\subsection*{Non-destructive 3D doping imaging}
\begin{figure}[!h]
\centering
    \makebox[\textwidth][c]{\includegraphics[width=170.6mm]{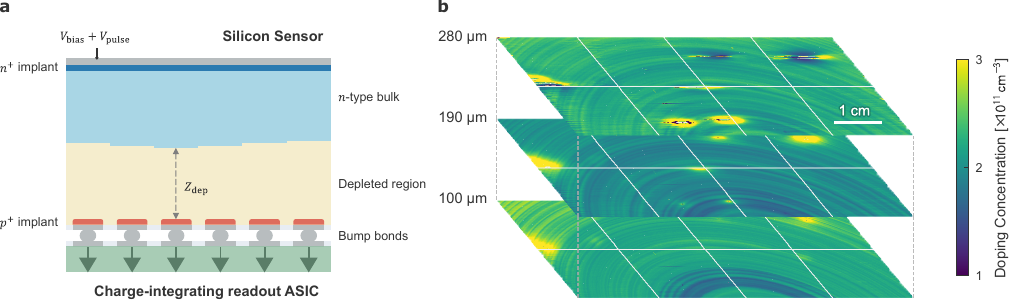}}
    \caption{ 
        $\vert$ \textbf{Non-destructive 3D doping imaging of operational silicon sensors.}
        \textbf{a}, Schematic of the doping profiling concept. A high-resistivity silicon sensor is bump-bonded to a charge-integrating Application-Specific Integrated Circuit (ASIC). Superimposing a synchronized voltage pulse ($V_{pulse}$) onto a sweeping DC bias ($V_{bias}$) allows depth-resolved charge injection and capacitance extraction across the entire pixel array.
        \textbf{b}, Reconstructed 3D doping concentration maps across the full sensor area, with $z = 0$ at the pixelated p$^{+}$ front-side electrode. Multi-depth slices (100~\textmu m, 190~\textmu m, and 280~\textmu m depths in this case) reveal both large-scale concentric ring patterns originating from the crystal growth process and localized doping anomalies, acquired simultaneously at 75~\textmu m lateral and 10~\textmu m depth sampling.
        Grid lines in between ASICs correspond to double-size pixels to accommodate the ASIC periphery, which are excluded from the mapping due to their intrinsically larger capacitance.
    }
    \label{fig1}
\end{figure}

To realize 3D doping imaging, we exploit the architecture of hybrid pixel detectors, in which a silicon sensor is bump-bonded to charge-integrating readout ASICs (Fig. \ref{fig1}a).
By sweeping a direct-current (DC) bias voltage applied to the sensor backside, the depletion region is progressively extended from the pixelated side into the bulk. 
A synchronized voltage pulse, superimposed onto this DC bias, capacitively injects a charge that, for a given pulse amplitude, is proportional to the local pixel capacitance and hence reports the depletion depth reached at that bias.
The readout ASICs capture this induced signal simultaneously across the entire pixel array, enabling the extraction of local pixel-to-backplane capacitance as a function of the applied bias voltage.

Rather than yielding a single, laterally averaged measurement, our method generates hundreds of thousands of independent $1/C^{2}-V$ curves across the entire pixel array.
By applying regional linear fitting, we map the doping concentration voxel by voxel. 
This yields a comprehensive, \textit{in-situ} 3D image of the bulk doping distribution across the sensor area (Fig. \ref{fig1}b).
The lateral resolution is intrinsically defined by the ASIC pixel pitch (75~\textmu m in this map, and scalable down to 25~\textmu m as demonstrated subsequently), while the depth resolution is $\sim$20~\textmu m, oversampled by a 10~\textmu m stepping.
Both limits are discussed in Methods.
By combining this micrometre-scale granularity with large-area coverage and a per-voxel reproducibility of $\sim 2\times10^{9}\ \text{cm}^{-3}$ at the concentrations probed here (Methods), this approach unlocks the wafer-scale structural discoveries and micro-defect analyses discussed below.

\subsection*{Wafer-scale doping structures}

\begin{figure*}[h]
\centering
    \makebox[\textwidth][c]{\includegraphics[width=157.4mm]{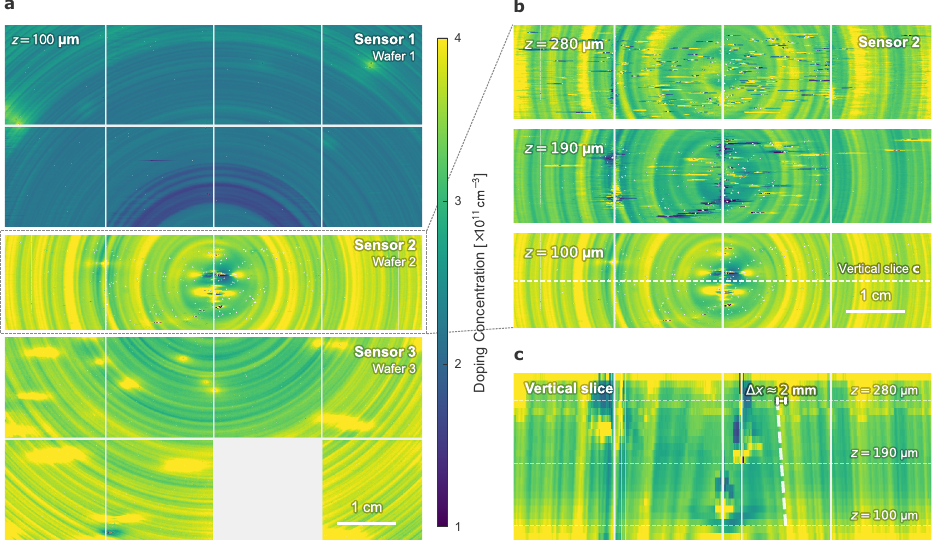}}
    \caption{$\vert$ \textbf{Wafer-scale doping imaging and heterogeneity.}
    \textbf{a}, Doping concentration maps extracted at $z = 100$~\textmu m for three large-area sensors, each map preserving the position and orientation of that sensor on its own wafer.
    The three sensors were diced from three wafers produced by one foundry.
    The maps are aligned on a common wafer centre, identifying the pattern as an inherent fingerprint of the float-zone crystal growth process. 
    Randomly scattered microscopic doping anomalies (isolated hotspots) are visible in addition. 
    The empty block in Sensor 3 corresponds to a defective ASIC, while the blank intersecting lines denote double-size pixels masked out due to their larger capacitance at the ASIC boundaries.
    \textbf{b}, Depth-resolved doping distributions of Sensor 2 at $z = 100$~\textmu m, $190$~\textmu m, and $280$~\textmu m. The concentric rings shrink dynamically as the depletion depth increases.
    \textbf{c}, A vertical cross-section (slice) of the bulk doping profile, quantitatively capturing the spatial dynamics of the macroscopic structures. The rings exhibit a significant lateral displacement ($\Delta x \approx 2~\text{mm}$) over a 180~\textmu m vertical span, indicating a highly complex, 3D non-uniform dopant incorporation. Localized anomalous structures extending vertically are also visible, shielding the electric field underneath. }
    \label{fig2}
\end{figure*}

We examined multiple sensors fabricated from different production batches to unveil the doping landscape at the wafer scale.
Figure~\ref{fig2}a presents the doping maps of three large-area sensors hybridized to 75~\textmu m-pixel pitch JUNGFRAU ASICs~\cite{Mozzanica_2016}, each diced from a different wafer of the same foundry (Sensors 1--3; Table~\ref{tab:sensors} in Methods). 
Each map preserves the position and orientation the sensor had on its own wafer.
Beyond revealing varying baseline doping levels of the sensors from wafer to wafer, Figure~\ref{fig2}a shows prominent concentric ring structures across all three sensors, revealing them to be inherent fingerprints of the float-zone crystal growth process.
While such macroscopic ring patterns have been empirically observed as underlying artifacts in the flat-field responses of an operational detector~\cite{Bergamaschi02112018} or the dark-field responses of under-depleted detectors, their volumetric physical origin has largely remained undocumented.
Here, we directly visualize these periodic fluctuations, which feature a peak-to-valley concentration ratio from $\sim$1.3 to $\sim$1.5:1. % 1.5:1 for the top sensor jf304
Whereas conventional techniques obscure these variations through lateral averaging or limited probing areas, our method captures their 3D topography within functional detectors.
As a physical consequence, these bulk doping gradients translate into a spatially correlated fluctuation in the depletion voltage, exhibiting a standard deviation of up to 10\% across the full sensor area (Extended Data Fig. \ref{fig:depletion}).

We further tracked the depth-dependent evolution of these concentric patterns, taking Sensor 2 as a representative example.
Figure \ref{fig2}b displays the doping distribution at three depths (100~\textmu m, 190~\textmu m, and 280~\textmu m), revealing that the rings shrink laterally as the depletion region extends deeper while the overall concentration level of the slice shifts with depth.
This spatial variation is quantitatively captured in the vertical cross-section (Fig. \ref{fig2}c) at the centre of the corresponding wafer.
Near the wafer centre, the rings shift laterally by $\Delta x \approx 2~\text{mm}$ over a 180~\textmu m span of sensor thickness, with the shift decreasing towards the sensor edge.
The iso-concentration surfaces therefore run only $\sim5^\circ$ from the wafer plane, i.e. $\sim85^\circ$ from the depth axis (note the strongly compressed vertical scale in Fig.~\ref{fig2}c). 

Beyond these macroscopic gradients, the 3D images also reveal discrete doping anomalies scattered across the sensor (Fig. \ref{fig2}a, b).
These intrinsic defects appear as vertically extended structures that shield underlying regions (Fig. \ref{fig2}c).
To quantify their impact on detector performance, the following section investigates these anomalies at a higher spatial resolution.

\subsection*{Microscopic doping anomalies and their impact on detector performance}

\begin{figure*}[h]
\centering
    \makebox[\textwidth][c]{\includegraphics[width=170mm]{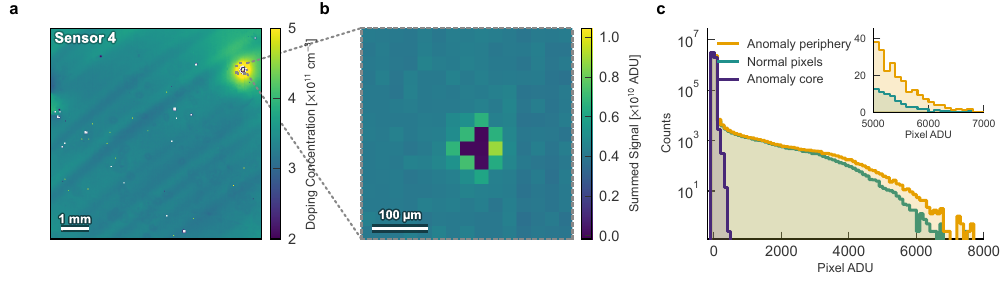}}
    \caption{$\vert$ \textbf{Microscopic doping anomalies and their impact on local charge collection and spectroscopic performance.}
    \textbf{a}, A high-resolution 2D doping concentration map at a depth of 200~\textmu m acquired using a 25 \textmu m-pitch MÖNCH ASIC, revealing a high-concentration anomaly embedded within the ring structures. 
    The apparent lateral extent of the high-doping ``halo" is inflated by local electric-field distortions, an extraction artifact detailed in the text.
    The profound electric field distortion creates a charge-depleted ``dark core'' and an artificially enhanced ``bright periphery'' where bordering pixels collect displaced charge carriers.
    \textbf{c}, Pixel-wise signal spectra (in ADU) for the three pixel populations indicated in b: the anomaly core, the bordering periphery, and normal pixels taken from the remainder of the map. 
    Compared to normal pixels, the anomaly core pixels exhibit a drastically suppressed response, whereas the anomaly periphery pixels show an excess of high-signal counts due to lateral charge redistribution, explicitly highlighted in the inset.}
    \label{fig3}
\end{figure*}

Focusing on these microscopic features, we utilized Sensor 4 (Table~\ref{tab:sensors}), a $1\times1~\text{cm}^2$ sensor bump-bonded to a MÖNCH ASIC~\cite{Ramilli_2017}, featuring a finer 25 \textmu m pixel pitch.
Figure~\ref{fig3}a presents the resulting high-resolution doping concentration map, which captures both the continuous rings and localized high-concentration doping anomalies. 
The lateral extent of these anomalies in the map is, however, overestimated: as shown below, local electric-field distortions break the planar-capacitance approximation and inflate the apparent footprint beyond its physical size.

To directly evaluate how such defects affect detector operation, we mapped the charge collection using flat-field illumination with 120 keV electrons. 
The resulting summed signal map (Fig. \ref{fig3}b) reveals a correlated disruption in the charge collection profile. 
The anomaly appears as a ``dark core'' surrounded by a bright periphery, originating from the lateral redistribution of charge carriers due to the distorted electric field.
Specifically, the elevated doping concentration within the anomaly hinders the downward growth of the depletion region and attracts the electric field lines from the surrounding pixels.
Furthermore, this field distortion violates the planar-capacitance assumption used in C-V profiling for nearby pixels.
This inflates the apparent lateral extent of the high-doping ``halo'' (Fig.~\ref{fig3}a), whereas the physical footprint of the defect in the charge collection map (Fig.~\ref{fig3}b) is confined to $\sim$100~\textmu m.
Within it, four core pixels lose 3.9 normal-pixel equivalents while their neighbours gain 7.8, the surplus indicating that the distorted field draws carriers from beyond the anomaly itself.

The redistributed charge collection is quantitatively reflected in the pixel-wise signal 
spectra (Fig.~\ref{fig3}c). 
As the track of a 120 keV electron spreads over a lateral scale comparable to the 25~\textmu m pitch, the charge of a single event is always shared among several pixels, so each spectrum is a continuum rather than a peak and the comparison is made between pixel populations. 
Relative to normal pixels taken from the remainder of the map, pixels bordering the anomaly accumulate more counts in the higher charge (higher Analog-to-Digital Units (ADU)) region, while pixels located within the anomaly core exhibit a heavily suppressed, yet non-zero, response, which excludes a disconnected readout channel as the origin.

Conventionally, malfunctioning pixels are attributed only to mechanical bump-bonding defects or readout electronics failures.
Our findings reveal that a substantial fraction of these failures instead originate from intrinsic doping anomalies within the silicon bulk.
Contiguous pixels whose response to the injected pulse deviates from the local median by more than 30\% are grouped into clusters.
A cluster is attributed to a bulk doping anomaly when it reproduces the core--periphery structure of Fig.~\ref{fig3}b, distinct from the signatures of the conventional failures reported in our previous work~\cite{XIE2026171227}.
Across two 25~\textmu m-pitch MÖNCH devices from two foundries, we find six such anomalies, an areal density of $\approx 3~\text{cm}^{-2}$, each spanning 10--40 contiguous pixels, so that approximately 0.04\% of all pixels are affected.
Given that state-of-the-art bump-bonding techniques achieve yields exceeding 99.95\%~\cite{Cartier:yn5011}, this intrinsic defect rate is comparable to the mechanical failure rate, representing a previously overlooked bottleneck in the pursuit of 100\% detector yield.

\section*{Conclusion}\label{sec3}

The ability to map 3D doping concentrations across large-area, functional detectors prompts a re-evaluation of sensor characterization.
By demonstrating \textit{in-situ} 3D doping imaging on assembled silicon sensors, this work simultaneously unravels the previously hidden doping landscape from macroscopic heterogeneities to microscopic anomalies.

Our findings reveal that the bulk doping of high-resistivity silicon is far more complex than the simplified assumption of uniform dopant distribution.
Macroscopically, the concentric doping rings arising from the crystal growth process result in a spatially dependent full-depletion voltage, with a relative standard deviation of up to 10\%.
To ensure uniform charge collection, detectors must therefore be operated at a bias voltage well exceeding the average full-depletion voltage.
Alongside this modulation within a single sensor, the baseline concentration itself differs from wafer to wafer, spanning $2$--$4\times10^{11}\ \rm{cm^{-3}}$ among the wafers sampled here, set by the ingot from which each wafer was cut rather than by the device process.
Microscopically, the localized doping anomalies act as a driver of pixel failure, distinct from and additive to conventional bonding and readout electronics issues. 
As bump bonding yields have been pushed beyond 99.95\%, this intrinsic contribution has become non-negligible, and cannot be removed by improving hybridization.

Despite its capabilities, this approach possesses specific operational constraints. 
The accuracy of the extracted doping concentration relies on the planar-capacitance approximation. 
Thus, robust profiling requires the depletion region to extend sufficiently deep into the bulk to minimize the influence of fringe fields. 
Similarly, as observed in the anomaly cores, electric field distortions violate this planar assumption, leading to extraction artifacts that inflate the anomaly footprint.
The applicability of this technique is inherently tied to the detector's architecture, requiring the sensor to deplete from its pixelated front side and to be coupled to charge-integrating readout ASICs.
Furthermore, as in any capacitance-based profiling method, the accessible resolution is bounded by the Debye length, over which free-carrier redistribution at the depletion edge smooths the underlying dopant distribution~\cite{Johnson1971,Kroemer1981}, limiting the depth resolution to $\sim$20~\textmu m here.

By exploring the 3D dopant landscape, this non-destructive 3D imaging technique serves as a powerful diagnostic tool for semiconductor foundries, offering comprehensive insights to optimize crystal growth and mitigate doping anomalies.
Beyond quality assurance, periodic profiling between irradiation steps leverages this rapid, \textit{in-situ} measurement capability to visualize radiation damage in extreme, high linear energy transfer environments~\cite{Lindstrom:1999mw}.
Moreover, it is adaptable to emerging sensor technologies, such as the avalanche breakdown in Low Gain Avalanche Detectors (LGADs)~\cite{Moffat_2018, BAHR2026171039, Zhao_2022} and the bulk electric field uniformity in Cadmium Zinc Telluride (CZT) sensors~\cite{10658160}, where the precise and homogeneous doping profile is critical for device performance and reliability.
Ultimately, this method provides a comprehensive tool to advance semiconductor sensors across diverse scientific disciplines. 

\section*{Methods}\label{sec11}

\subsection*{Silicon sensors and readout ASICs}
The experimental validation of our 3D doping imaging technique was performed using hybrid pixel detectors ({Fig. \ref{fig:setup}a}), which consist of a high-resistivity silicon sensor bump-bonded to charge-integrating readout ASICs. 
The sensing volumes are composed of n-type float-zone silicon with $\langle 111 \rangle$ surface orientation, featuring a continuous $n^+$ backside electrode and a pixelated $p^+$ electrode on the front side formed via boron implantation.
Five sensors from five different wafers were evaluated; their provenance, geometry and readout configuration are summarized in Table~\ref{tab:sensors}.
Charge-integrating readout ASICs employed in this study include the JUNGFRAU~\cite{Mozzanica_2016} and MÖNCH~\cite{Ramilli_2017} chips, designed in-house and fabricated using UMC 110 nm CMOS technology.
Prior to hybridization, the sensor wafers received additional lithography steps to form the bump-bonding interface: openings in the front-side passivation exposing each p$^{+}$ implant, followed by deposition of an under-bump metallization stack and an indium bump on every pixel.
These front-side processes leave the bulk doping distribution unaffected.

\begin{table}[h]
\centering
\caption{\textbf{Overview of the silicon sensors characterized in this work.}
All sensors are n-type $\langle 111 \rangle$ float-zone silicon with a continuous backside electrode and a boron-implanted pixelated front side.
Sensors 1--4 were produced by Foundry A on four different wafers; Sensor 5 originates from Foundry B.}
\label{tab:sensors}
\begin{tabular}{@{}llcccl@{}}
\toprule
Sensor & Foundry / wafer & Thickness (\textmu m) & Readout ASIC & Pitch (\textmu m) & Active area (cm$^{2}$) \\
\midrule
1 & A / W1 & 320 & $8\times$ JUNGFRAU & 75 & $7.7\times3.9$ \\
2 & A / W2 & 320 & $4\times$ JUNGFRAU & 75 & $7.7\times1.9$ \\
3 & A / W3 & 320 & $8\times$ JUNGFRAU & 75 & $7.7\times3.9$ \\
4 & A / W4 & 320 & MÖNCH            & 25 & $1\times1$ \\
5 & B / W5 & 300 & MÖNCH            & 25 & $1\times1$ \\
\bottomrule
\end{tabular}
\end{table}

For macroscopic wafer-scale imaging, sensors 1--3 with large active areas of $7.7 \times 3.9\ \text{cm}^{2}$ and $7.7 \times 1.9\ \text{cm}^{2}$ were hybridized to an array of eight and four JUNGFRAU ASICs, defining a lateral spatial resolution of 75 \textmu m.
In between the ASICs, one row and column of pixels are doubled to $150\ \text{\textmu m}$ to accommodate the ASIC periphery, which are excluded from the doping mapping due to their larger capacitance.
The outermost 15 pixel rows and columns of every sensor are also excluded from the analysis, as their capacitance carries an additional contribution from the adjacent guard-ring structure.

Conversely, for high-resolution microscopic defect analysis, smaller sensors ($1 \times 1\ \text{cm}^{2}$ active area) were bump-bonded to single MÖNCH ASICs, providing a finer 25 \textmu m pixel pitch. 
Two such devices were characterized: Sensor 4 from Foundry A, used for the microscopic anomaly analysis in Fig.~\ref{fig3}, and Sensor 5 from Foundry B, included in Fig.~\ref{fig:extraction} as a second, independently sourced device.

\subsection*{Synchronized backside pulsing and data acquisition}

\begin{figure}[h]
\centering
    \makebox[\textwidth][c]{\includegraphics[width=172 mm]{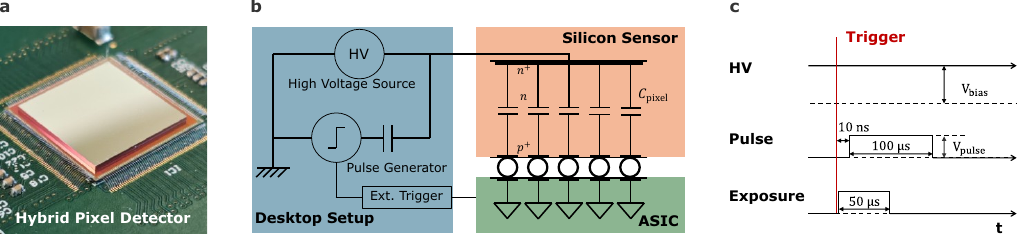}}
    
    \caption{$\vert$ \textbf{Experimental setup and synchronized data acquisition scheme.} 
    \textbf{a}, Photograph of a representative hybrid pixel detector, consisting of a silicon sensor bump-bonded to a charge-integrating readout ASIC. 
    \textbf{b}, Schematic of the desktop-based testing configuration. A swept DC bias voltage ($V_{\rm{bias}}$) is applied to establish controlled depletion depths, onto which a customized voltage pulse is capacitively superimposed at the sensor backside to inject charge and probe the pixel capacitance ($C_{\rm{pixel}}$).
    \textbf{c}, Timing logic for data acquisition. An external transistor-transistor logic (TTL) trigger synchronizes the detector exposure window (50 \textmu s) with the voltage pulse (100 \textmu s). A 10 ns delay ensures that only the rising edge of the pulse is integrated by the ASIC preamplifier.}
    \label{fig:setup}
\end{figure}

The desktop-based testing system utilizes a Keithley 2410~\cite{Keithley2410} to supply a DC bias voltage ($V_{\rm{bias}}$), which controls the progressive extension of the depletion region. 
Onto this DC bias, we superimpose voltage pulses generated by an Agilent 33250A~\cite{33250A} function generator, amplified by a custom-built amplifier (gain $\approx 4.4$), and capacitively coupled to the sensor backside ({Fig. \ref{fig:setup}b}).
An external 1 kHz TTL trigger synchronizes the detector exposure with the pulse generation under a 10 ns delay ({Fig. \ref{fig:setup}c}). 
The pulse duration is configured to 100 \textmu s, which exceeds the 50 \textmu s detector exposure window but remains shorter than the 600 \textmu s readout period. 
This timing scheme guarantees that only the rising edge of the voltage pulse is integrated by the charge-sensitive preamplifier, avoiding signal cancellation from the falling edge~\cite{Mezza_2016}.  

During data acquisition, the detector is maintained at a constant temperature of 293 K using a liquid cooling system.
To extract pixel-wise capacitance-voltage profiles, $V_{\rm{bias}}$ is systematically scanned from near 0 V to above the full depletion voltage.
The bias increments follow a square-root distribution to approximate linear sampling intervals of the depletion depth ($z_{\rm{dep}}$).
Concurrently, the pulse amplitude ($V_{\rm{pulse}}$) is scaled proportionally to $\sqrt{V_{\rm{bias}}}$. 
This strategy maintains an optimal signal level and prevents electronic saturation at lower bias voltages where the effective pixel capacitance peaks. 
The scan comprises approximately 500 bias steps, with the voltage increment growing from $\sim$5~mV to $\sim$100~mV near full depletion.
This corresponds to a spacing in depletion depth of about 1~\textmu m, so that each 10~\textmu m interval used for the local linear fitting is populated by about 10 independent measurements.
At each voltage step, 1,000 frames are acquired both with and without the pulse to enable pedestal subtraction and statistical noise suppression.
A complete scan takes approximately 5~h per detector, set mainly by the settling between steps.

The amplitude of the voltage pulse delivered to the sensor backside was calibrated \textit{in-situ} with an oscilloscope connected in parallel to the biased detector. 
The amplifier gain $G$ was found to be independent of the pulse amplitude to within 1\% over the range employed, but to vary weakly and linearly with the applied bias, $G(V_{\rm{bias}}) = 4.39 + 2.3\times10^{-3}\cdot\,V_{\rm{bias}}$ for the MÖNCH 
setup and $G(V_{\rm{bias}}) = 4.45 + 2.1\times10^{-3}\cdot V_{\rm{bias}}$ for the
JUNGFRAU setup, both calibrated up to 100~V.
This dependence is applied point-by-point in the extraction. 

\subsection*{3D doping extraction}
Following pedestal subtraction, the raw detector output is first calibrated to energy using our previously established pixel-wise look-up table~\cite{XIE2026171227}, and then converted into the injected charge ($Q_{\rm{pixel}}$) using the electron-hole pair generation energy of $3.62\ \text{eV}$ for silicon~\cite{Eehpair_3_62eV}.
Assuming a planar-capacitance model where $C_{\rm{pixel}} = \epsilon_{\rm{Si}}A_{\rm{pixel}}/z_{\rm{dep}}$, the local depletion depth, $z_{\rm{dep}}$, for each individual pixel is formulated as:$$z_{\rm{dep}} = \frac{\epsilon_{\rm{Si}} V_{\rm{pulse}} A_{\rm{pixel}}}{Q_{\rm{pixel}}}$$where $\epsilon_{\rm{Si}}$ is the permittivity of silicon and $A_{\rm{pixel}}$ represents the pixel area.
The depth coordinate is measured from the pixelated p$^{+}$ front-side electrode throughout this work, which defines $z = 0$.

Subsequently, the localized bulk doping concentration, $n(x,y,z)$, is extracted from the voltage derivative of the squared inverse capacitance, following the principle of C-V profiling:
$$n(x,y,z) = \frac{2}{q_{0} \epsilon_{\rm{Si}} A_{\rm{pixel}}^{2}} \left[ \frac{\mathrm{d}(1/C_{\rm{pixel}}^{2})}{\mathrm{d}V_{\rm{bias}}} \right]^{-1}$$ where $q_{0}$ is the elementary charge.

\begin{figure}[h]
\centering
    \makebox[\textwidth][c]{\includegraphics[width=140 mm]{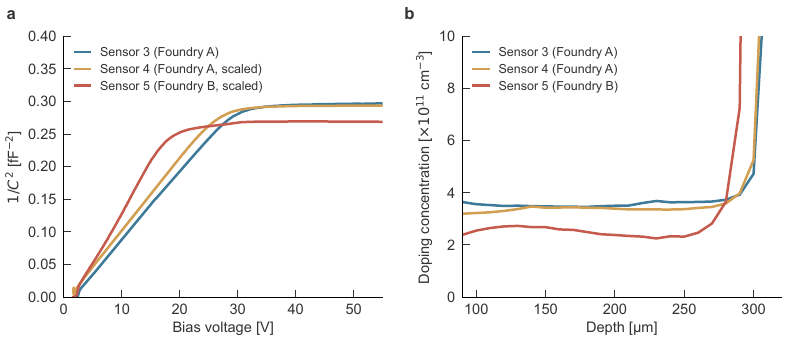}}
    \caption{$\vert$ \textbf{C--V curves and extracted depth-resolved doping profiles.} 
        \textbf{a}, Median $1/C_{\rm{pixel}}^{2}-V_{\rm{bias}}$ curves for the evaluated sensors. 
        Sensor 3 is shown as measured; the MÖNCH-based Sensors 4 and 5 are scaled to the JUNGFRAU pixel area for direct comparison.
        \textbf{b}, Corresponding median depth-dependent doping concentration profiles. 
        The rise beyond $\sim280$~\textmu m marks the onset of the highly doped n$^{+}$ backside implant.
    }
    \label{fig:extraction}
\end{figure}

{Fig. \ref{fig:extraction}a} depicts the pixel-wise median $1/C_{\rm{pixel}}^{2}-V_{\rm{bias}}$ curves of Sensors 3--5 (Table~\ref{tab:sensors}).
Sensor 3 is read out by JUNGFRAU ASICs and is shown as measured, whereas the curves of Sensors 4 and 5, acquired with the finer 25~\textmu m MÖNCH pitch, are scaled to the JUNGFRAU pixel area to allow a direct comparison on a common axis.
The non-linearity of the curves reflects longitudinal fluctuations in doping concentration.
We evaluate the derivative through localized linear fitting at discrete depth intervals of 10 \textmu m, each containing about 10 bias points, treating the intercept in each local fit as an unconstrained free parameter.
The intrinsic depth resolution of any C-V-based profiling is bounded by the Debye length, $L_{\rm{D}} = \sqrt{\epsilon_{\rm{Si}} k_{\rm{B}} T / q_{0}^{2} n}$, over which free-carrier redistribution at the depletion edge smooths the underlying dopant distribution, so that structures narrower than a few $L_{\rm{D}}$ cannot be resolved~\cite{Johnson1971,Kroemer1981}.
For the bulk concentrations probed here, $L_{\rm{D}}$ ranges from 8.0 to 6.8~\textmu m at $2.5$ and $3.5\times10^{11}\ \rm{cm^{-3}}$, respectively, limiting the depth resolution to $\sim$20~\textmu m, so that the 10~\textmu m sampling step already oversamples the achievable resolution.
All structures reported in this work are considerably larger, and no deconvolution correction is applied.

For each 10~\textmu m depth interval, the pixel-wise doping values are aggregated into median profiles for the sensors (Fig.~\ref{fig:extraction}b).
Sensor~3 and Sensor~4, diced from different wafers of Foundry A, exhibit similar doping levels, whereas Sensor~5 shows a lower baseline concentration.
Since the bulk doping is fixed at crystal growth rather than during device processing, this offset reflects the ingot each wafer was cut from, as does the wafer-to-wafer spread observed within Foundry A itself.

The reproducibility of the extraction was assessed by repeating the full measurement and the same analysis chain on Sensor 5 in two sessions separated by several days.
Over the depth range 90--280~\textmu m, the relative difference between the two independent measurements is centred at 0.30\% with a standard deviation of 0.89\%  (Extended Data Fig.~\ref{fig:doping_reproducibility}), confirming that the absolute scale is stable between sessions.
As this distribution refers to the difference of two independent measurements, the per-voxel reproducibility of a single measurement is $0.89\%/\sqrt{2} = 0.63\%$, i.e. $\sim2\times10^{9}\ \rm{cm^{-3}}$ at the concentrations probed here.

As an independent check of the absolute scale, the concentrations extracted here can be compared with the resistivity reported by Foundry B.
Given the standard resistivity--dopant density conversion $n = 1/(q_{0}\mu_{n}\rho)$ with the lattice-limited electron mobility $\mu_{n} = 1450\ \rm{cm^{2}V^{-1}s^{-1}}$~\cite{Jacoboni1977}, the reported resistivity of 15.3--17.8 k$\Omega\,$cm corresponds to $2.4$--$2.8\times10^{11}\ \rm{cm^{-3}}$, bracketing the $\sim2.5\times10^{11}\ \rm{cm^{-3}}$ obtained for Sensor 5 (Fig.~\ref{fig:extraction}b).

\subsection*{Pixel-wise depletion voltage extraction}

For each pixel, $V_{\rm{dep}}$ is determined by calculating the intersection of two linear fits applied to its $1/C_{\rm{pixel}}^{2}-V_{\rm{bias}}$ curve: one characterizing the partially depleted bulk in the low-bias regime, and the other defining the fully depleted plateau in the high-bias regime. 
Pixels exhibiting poor fit quality, typically due to defective readout channels, are excluded from the macroscopic maps.

The resulting wafer-scale depletion voltage distribution ({Extended Data Fig. \ref{fig:depletion}a}) mirrors the concentric ring patterns observed in the doping maps ({Fig. \ref{fig2}a}).
This spatial correlation confirms that the dopant heterogeneity influences the electrical characteristics of the sensor.
Furthermore, statistical analysis of these spatial distributions (Extended Data Fig.~\ref{fig:depletion}b) reveals systematic differences between wafers of the same foundry. 
Sensor~1 exhibits the lowest mean $V_{\rm{dep}}$ (18.4~V), whereas Sensors~2 and 3 show higher values (28.4 and 29.1~V). 
The corresponding standard deviations, computed over all pixels passing the fit-quality selection and without any outlier rejection, are 1.6, 1.8 and 2.8~V, i.e. 6--10\% of the respective mean.
The parallel-plate relation $V_{\rm{dep}} = q_{0} n d^{2} / 2\epsilon_{\rm{Si}}$ converts these baselines into bulk concentrations of $2.3$, $3.6$ and $3.7\times10^{11}\ \rm{cm^{-3}}$, consistent with the doping levels mapped directly in Fig.~\ref{fig2}a and, for Sensor~3, with the $\sim3.5\times10^{11}\ \rm{cm^{-3}}$ extracted independently in Fig.~\ref{fig:extraction}b.

\subsection*{Flat-field charge collection mapping}

Sensor 4 was evaluated under flat-field illumination using 120 keV electrons incident on the backside.
With a continuous-slowing-down range of a few tens of micrometres in silicon, the electrons deposit their energy within a thin layer at the entrance surface, so that essentially all signal carriers drift across the full sensor thickness and traverse the depth range of the anomaly.
Data were acquired at low occupancy to prevent saturation and with sufficient statistics for a pixel-wise analysis.
The pedestal-subtracted pixel responses were analysed in ADU, as the calibration was locally not applicable due to the distorted electric field in the anomaly region.
By integrating the total collected signal per pixel, we mapped the spatial footprint of the lateral charge redistribution (Fig.~\ref{fig3}b).
Signal deviations from the normal-pixel level, taken as the median over an annulus 16--30 pixels from the core, were summed over square regions of increasing half-width.
The integral saturates beyond a half-width of four pixels, and is quoted in units of the normal-pixel level.
Correspondingly, individual pixel spectra were extracted to evaluate the degree of signal suppression and excess charge collection (Fig.~\ref{fig3}c).

\section*{Data availability}
The data that support the findings of this study are available in the Zenodo
repository at \url{https://doi.org/10.5281/zenodo.22127229}.

\section*{Code availability}
The analysis chain used to reconstruct the doping maps and to generate the
figures of this study is available in the Zenodo repository at
\url{https://doi.org/10.5281/zenodo.22127883}.

% \bmhead{Supplementary information}

% Supplementary Video 1: depth scan of the reconstructed doping concentration map of Sensor~2.

\bmhead{Acknowledgements}

The authors sincerely appreciate the helpful technical discussions with Dr. Gian-Franco Dalla Betta and Dr. Michael Moll regarding the interpretation of the results and their implications for sensor development.

We acknowledge the usage of the instrumentation provided by the Electron Microscopy Facility at PSI and we thank Dr. Elisabeth Müller and Dr. Emiliya Poghosyan for their support with electron microscopy data acquisition.

\bmhead{Funding}
This work was supported by internal funding of the Paul Scherrer Institute. 

\bmhead{Author contributions}
X.X. designed and performed the experiments, developed the analysis chain, interpreted the results and wrote the manuscript. 
J.Z. conceived the study. 
E.F. developed the data-analysis library used for the raw-data processing.
D.M. developed the backside charge-injection concept underlying the pulsing scheme. 
M.C. provided the silicon sensors.
R.D. and A.M. designed the MÖNCH and JUNGFRAU ASICs.
V.H., J.M. and S.S. contributed to the discussion of the results and to the critical revision of the manuscript.
A.B. and B.S. supervised the project and provided resources.
All authors read and approved the final version of the manuscript.

\bmhead{Competing interests}
The authors declare no competing interests.
\backmatter

% \noindent
% If any of the sections are not relevant to your manuscript, please include the heading and write `Not applicable' for that section. 

% \bigskip

% Springer journals and proceedings: \url{https://www.springer.com/gp/editorial-policies}

% Nature Portfolio journals: \url{https://www.nature.com/nature-research/editorial-policies}

% \textit{Scientific Reports}: \url{https://www.nature.com/srep/journal-policies/editorial-policies}

% BMC journals: \url{https://www.biomedcentral.com/getpublished/editorial-policies}
% \end{flushleft}

\begin{appendices}

\section*{Extended Data}\label{sec:supplementary}

\setcounter{figure}{0}
\renewcommand{\theHfigure}{SF\arabic{figure}}   % hyperref
\renewcommand{\figurename}{Extended Data Fig.}

\begin{figure}[H]
\centering
    \makebox[\textwidth][c]{\includegraphics[width=135 mm]{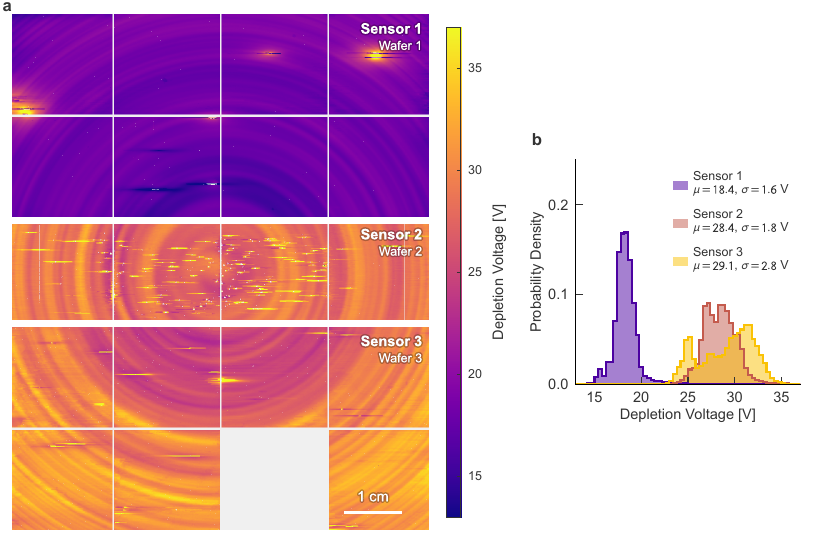}}
    
    \caption{\textbf{$\vert$ Wafer-scale mapping and statistics of the depletion voltage.} 
        \textbf{a}, Depletion voltage ($V_{\rm{dep}}$) arranged as in Fig.~\ref{fig2}a. The spatial fluctuations in $V_{\rm{dep}}$ strongly correlate with the concentric doping ring patterns (Fig.~\ref{fig2}a).
        \textbf{b}, Probability density distributions of the depletion voltages for Sensor 1, 2, and 3. 
        % The statistical profiles reveal systematic variations in the baseline depletion voltage and uniformity across different wafers.
        }
    \label{fig:depletion}
\end{figure}

\begin{figure}[H]
\centering
    \includegraphics[width=72 mm]{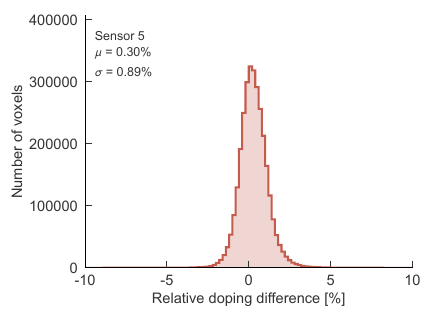}
    \caption{\textbf{$\vert$ Reproducibility of the 3D doping extraction.}
    Distribution of the relative difference in extracted doping concentration between two independent measurements of Sensor 5 bump-bonded to a MÖNCH ASIC with 160,000 pixels, acquired several days apart and processed with an identical analysis chain. 
    Each entry corresponds to one voxel within the depth range 90--280~\textmu m; voxels with extracted concentrations outside $1$--$5\times10^{11}\ \rm{cm^{-3}}$, or with a relative difference exceeding 10\%, are excluded (0.5\% of the total). 
    The distribution is centred at $\mu = 0.30\%$ with a standard deviation of $\sigma = 0.89\%$, showing that the absolute scale is stable at the sub-percent level between sessions. 
    As the histogram represents the difference of two independent measurements, the reproducibility of a single measurement is $\sigma/\sqrt{2} = 0.63\%$ per voxel. 
    }
    \label{fig:doping_reproducibility} 
\end{figure}

% An appendix contains supplementary information that is not an essential part of the text itself but which may be helpful in providing a more comprehensive understanding of the research problem or it is information that is too cumbersome to be included in the body of the paper.




\end{appendices}


% \bibliography{references}% common bib file


\end{document}